\documentclass[12pt,twoside]{article}   %For printing on two sides of page
\usepackage[super,sort,comma]{natbib}
\usepackage{fancyhdr}		%Gives headers and footers defined below
\usepackage[section]{placeins}   %
\usepackage{graphicx}
\usepackage{algorithmic}
\usepackage{algorithm}
\usepackage{amsmath,amssymb,amsfonts}
\usepackage{makecell}
\usepackage{capt-of}

\makeatletter \renewcommand\@biblabel[1]{$^{#1}$} \makeatother
\newcommand{\cen}[1]{\begin{center} #1 \end{center}}

\usepackage[mathlines]{lineno}
\usepackage{hyperref}
\hypersetup{ colorlinks,
    citecolor=blue,
    filecolor=blue,
    linkcolor=blue,
    urlcolor=blue
}

\usepackage{xcolor}
\usepackage{pifont}
\definecolor{gray}{rgb}{0.6,0.6,0.6}
\definecolor{red}{rgb}{0.85,0,0}
\definecolor{green}{rgb}{0,0.85,0}
\definecolor{blue}{rgb}{0,0,0.85}
\definecolor{beige}{rgb}{0.92,0.87,0.78}
\usepackage[all]{hypcap}    %causes link to figures to go to figure, not caption
\begin{document}

\cen{\sf {\Large {\bfseries Limited Parameter Denoising for Low-dose X-ray Computed Tomography Using Deep Reinforcement Learning } \\  
\vspace*{10mm}
Mayank Patwari$^{1, 2}$, Ralf Gutjahr$^{2}$, Rainer Raupach$^{2}$, Andreas Maier$^{1}$} \\
1. Pattern Recognition Lab, Friedrich-Alexander Universität Erlangen-Nürnberg, 91058 Erlangen, Germany \\
2. CT Concepts, Siemens Healthineers AG, 91301 Forchheim, Germany
\vspace{5mm}\\
Version typeset \today\\
}

\pagenumbering{roman}
\setcounter{page}{1}
\pagestyle{plain}
Mayank Patwari, email: mayank.patwari@fau.de \\
% note, probably best not to use a student's e-mail as it won't be valid for
% very long.

\begin{abstract}
\noindent {\bf Purpose:} The use of deep learning has  successfully solved several problems in the field of medical imaging. Deep learning has been applied to the CT denoising problem successfully. However, the use of deep learning requires large amounts of data to train deep convolutional networks (CNNs). Moreover, due to large parameter count, such deep CNNs may cause unexpected results. In this study, we introduce a novel CT denoising framework, which has interpretable behaviour, and provides useful results with limited data.\\
{\bf Methods:} We employ bilateral filtering in both the projection and volume domains to remove noise. To account for non-stationary noise, we tune the $\sigma$ parameters of the volume for every projection view, and for every volume pixel. The tuning is carried out by two deep CNNs. Due to impracticality of labelling, the two deep CNNs are trained via a Deep-Q reinforcement learning task. The reward for the task is generated by using a custom reward function represented by a neural network. Our experiments were carried out on abdominal scans for the Mayo Clinic TCIA dataset, and the AAPM Low Dose CT Grand Challenge.\\
{\bf Results:} Our denoising framework has excellent denoising performance increasing the PSNR from 28.53 to 28.93, and increasing the SSIM from 0.8952 to 0.9204. We outperform several state-of-the-art deep CNNs, which have several orders of magnitude higher number of parameters ($p$-value (PSNR) = 0.000, $p$-value (SSIM) = 0.000). Our method does not introduce any blurring, which is introduced by MSE loss based methods, or any deep learning artifacts, which are introduced by WGAN based models. Our ablation studies show that parameter tuning and using our reward network results in the best possible results. \\
{\bf Conclusions:} We present a novel CT denoising framework, which focuses on interpretability to deliver good denoising performance, especially with limited data. Our method outperforms state-of-the-art deep neural networks. Future work will be focused on accelerating our method, and generalizing to different geometries and body parts. \\

\end{abstract}
%\note{This is a sample note.}

\newpage     %may or may not be needed

%The table of contents is for drafting and refereeing purposes only. Note
%that all links to references, tables and figures can be clicked on and
%returned to calling point using cmd[ on a Mac using Preview or some
%equivalent on PCs (see View - go to on whatever reader).
%\tableofcontents

\newpage

\setlength{\baselineskip}{0.7cm}      %double spacing		
\setlength{\belowcaptionskip}{0.7cm}

\pagenumbering{arabic}
\setcounter{page}{1}
\pagestyle{fancy}

\section{Introduction}

The increasing use of CT in medical diagnosis and intervention comes with several advantages and disadvantages. A major disadvantage is the use of damaging radiation for scanning \cite{Monga2007,Remedios2020,Yu2009,Kachelriess2020}. To reduce patient harm, the radiation dose of CT is kept as low as possible \cite{Lee2020,Kalra2004,Kalra2005,Yu2010}, however, this results in noise in the collected projections and subsequently in the reconstructed volume \cite{Oppelt2005a}. Noise reduces the clinical value of the CT scan.

Noise removal in CT is usually performed in conjunction with the reconstruction process \cite{He2019,Manduca2009,Li2014,Wu2017a,Maier2011,Lorch2015}. Iterative CT reconstruction \cite{Geyer2015,Gilbert1972,Kaczmarz1937,Beister2012}, which is currently offered on most clinical scanners, uses noise removal filters in the projection and reconstructed volume domains to remove noise, while using the iterative nature of the process to resolve geometric artifacts. Examples of iterative reconstruction algorithms include Siemens ADMIRE \cite{Ramirez-Giraldo2015} and Canon AIDR \cite{Angel2012}. Both of these have been shown to reduce noise and improve image quality.

Deep learning approaches have been successfully applied to the CT denoising problem \cite{Maier2015,Patwari2020,Liu2018}. Most deep learning approaches for CT denoising are formulated in the form of image translation tasks\cite{Wolterink2017,Zhang2017a,Yang2018,Li2020,Fan2019,Patwari2020b,Shan2018,Chen2017}. CNNs learn a mapping from a noisy CT volume to a clean CT volume. Some networks also attempt to denoise CT volumes iteratively \cite{Liu2019,Wu2017a,Shen2018,Shan2019}, for better control of the denoising processes. A few methods attempt to model the reconstruction process or parameters \cite{He2019,Wu2017a,Syben2017,Li2018,Zhang2020a}. Yin et al. \cite{Yin2019} applies separate deep neural networks, trained separately, for denoising in the projection and volume domains. Deep reinforcement learning has been applied to tune pixelwise parameters for solving the reconstruction \cite{Shen2018} and denoising \cite{Patwari2020a} problems. 

While deep learning based methods have shown excellent performance, most deep learning methods have several thousand trainable parameters, which affects the interpretability of the solution, and may result in unexpected behaviour. Moreover, deep learning methods require a large amount of data to learn the image translation from the low dose to standard dose CT.

\begin{figure}[t]
	\includegraphics[width=\linewidth]{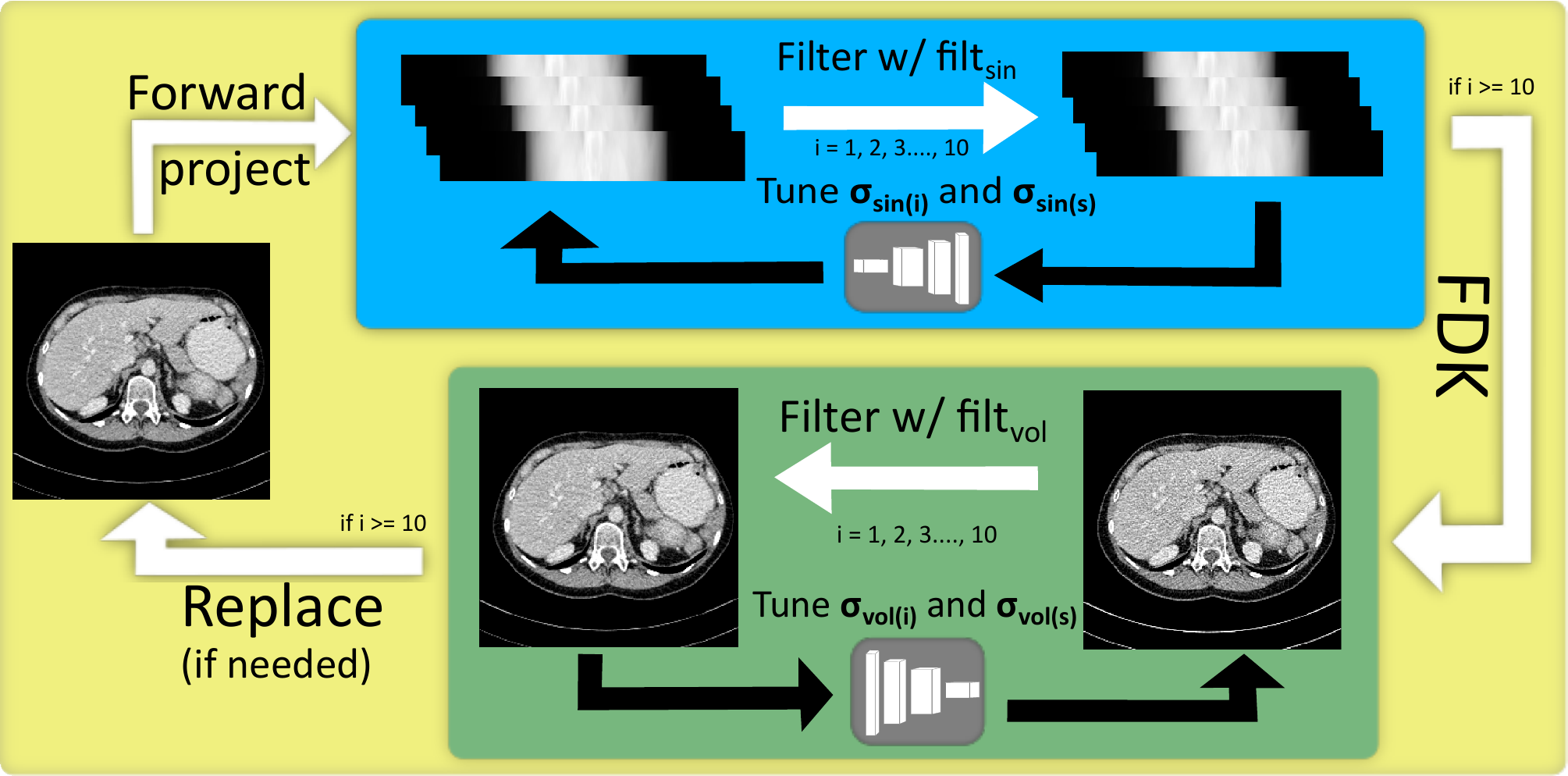}
	\caption{A schematic explaining our filtering scheme. We simulate a forward projection, tune parameters and apply $filt_{sin}$, reconstruct using FDK, tune parameters and apply $filt_{vol}$. Replacement and iteration is possible, although not required in our study.}
	\label{denoisScheme}
\end{figure}

In this study, we aim to combine the prior knowledge of the CT reconstruction/denoising process, with the power of deep learning based methods, to simultaneously denoise in the projection and volume domains. We assume that incorporating prior knowledge will lead to better results \cite{Maier2019}. Our noise removal filters are modelled by bilateral filters to filter in both the projection ($filt_{sin}$) and volume domain ($filt_{vol}$) (see Fig.\ref{denoisScheme}). Each bilateral filter has two $\sigma$ parameters, which control the strength of the filtering and smoothing. 

We introduce two CNNs $NET_{sin}$ and $NET_{vol}$, to tune the $\sigma$ parameters in each domain. Each network chooses which parameter needs to be optimized, and then an action to change the value of the parameter. $filt_{sin}$ has different $\sigma$ values for each projection view, while $filt_{vol}$ has different $\sigma$ values for each pixel. We update the values of $NET_{sin}$ and $NET_{vol}$ using a Deep Q learning scheme \cite{Mnih2015}, which has been previously employed in CT denoising \cite{Patwari2020a,Shen2018}. Both networks are trained independently of one another, thereby decoupling the denoising process from the reconstruction algorithm. This enables the use of different reconstruction algorithms to solve geometric artifacts. The reward used for Deep Q learning is estimated using a reward network $NET_{rew}$. 

We demonstrate that our denoising framework results in better structure preserving noise removal compared to classical image-to-image deep learning methods. Since we use a reward network, paired ground-truth volumes are not required for the training process. Additionally, we also demonstrate that, in contrast to state-of-the-art methods, we are able to achieve good denoising performance with only 10 volumes as part of our training dataset. We demonstrate this on a subset of 50 patients with abdominal scans taken from the Mayo Clinic TCIA dataset \cite{Mccollough}.

\section{Materials and Methods}

\subsection{CT Denoising Framework}
\label{denFram}
Our CT denoising framework is based on the use of the bilateral filter in both the projection ($filt_{sin}$) and volume ($filt_{vol}$) domains.  We will review the bilateral filter, to make our work self-contained. 

The bilateral filter is a filter which accounts for the differences in local CT values, as well as the difference in local coordinates. It is represented by the following equation:
\begin{equation}
	I_f(x) = \frac{\sum_{o \epsilon N(x)} I_n(o) G_{\sigma_s}(x - o) G_{\sigma_i}(I_n(x) - I_n(o))  }{\sum_{o \epsilon N(x) } G_{\sigma_s}(x - o)  G_{\sigma_i}(I_n(x) - I_n(o))  }
\end{equation}

\noindent where $I_n$ is the noisy image, $I_f$ is the filtered image, $x$ is the spatial coordinate, and $G_{\sigma}$ is the Gaussian operator. The Gaussian operator is defined by the following equation:
\begin{equation}
	G_{\sigma}(x) = \frac{e^{-x^2 / 2\sigma^2} }{2\sigma^2}
\end{equation}

\noindent where $\sigma$ is a hand tuned parameter, which determines the strength of filtering. There are two $\sigma$ parameters in a bilateral filter, $\sigma_i$, which controls the strength of the difference of \textbf{i}ntensities (CT values), and $\sigma_s$, which controls the strength of the difference of \textbf{s}patial coordinates. Both $filt_{sin}$ and $filt_{vol}$ have two $\sigma$ parameters, resulting in a total of four parameters to control the noise removal process. $filt_{sin}$ has a 5 $\times$ 5 neighborhood, while $filt_{vol}$ has a 5 $\times$ 5 $\times$ 5 neighborhood.

We first apply $filt_{sin}$ to our projections. After filtering, we reconstruct the filtered projections. Since we are using cone-beam data, we reconstruct using the FDK algorithm \cite{Feldkamp1984}. To account for residual noise amplified or magnified during the reconstruction process, we apply $filt_{vol}$ to our reconstructed volume to remove any leftover noise (Fig. \ref{denoisScheme}). In this way, the noise removal process is decoupled from the actual reconstruction algorithm, allowing use of an arbitrary reconstruction algorithm.

\subsection{Parameter Tuning Method}

Due to the non - stationary nature of CT noise, the best image quality would be achieved by finding the ideal $\sigma$ values for each pixel in the noisy image or projection. A human operator would do this by choosing a single $\sigma$ parameter, tuning the value globally, and observing how it affects the resulted image. If this satisfies the operator, they would leave the image as it is. Otherwise, they would change the strength, or choose another parameter, and use the new parameter and strength combination to filter the image. Since repeatedly filtering an image would cause blurring, the new tuned parameter would be used to filter the original noisy image. However, a volume of 520 $\times$ 256 $\times$ 256, a realistically sized volume for a reconstructed CT scan, would have over 34 million pixels. Adjusting the $\sigma$ parameters by hand for these many pixels would be impractical. 

Therefore, we develop an automatic parameter tuning method which can help us to find optimal parameter values. We make some assumptions to simplify our problem. We assume that, since the noise in the projection domain follows a Poisson 
distribution, a global $\sigma$ value for each projection view should be sufficient to remove the noise. There are two reasons why we make this assumption. The first is that, the data we are using is scanned by CT systems which use a bow-tie filter. This largely means that the absorption across the projection view is uniform after filtering, therefore the non-stationarity of the noise is reduced, as opposed to CT systems without bow-tie filters \cite{Borsdorf2008a,Lu2015,Manhart2014,Maier2015}.

\vspace{0.2cm}
\begin{algorithm}
	\caption{CT Denoising Framework}
	\begin{algorithmic}[1]
		\REQUIRE $filt_{sin}, filt_{vol}, NET_{sin}, NET_{vol}, vol_{noisy}$
		\STATE $proj = forward~project(vol_{noisy})$
		\STATE $proj_{backup} = proj$
		\FOR {$i$ in 1, 2, ..., 10}
		\STATE tune $filt_{sin}~\sigma$ params $\rightarrow NET_{sin}(proj_{backup})$
		\STATE $proj_{filt} = filt_{sin}(proj)$
		\STATE $proj_{backup} = proj_{filt}$
		\ENDFOR
		\STATE $vol = FDK(proj_{filt})$
		\STATE $vol_{backup} = vol$
		\FOR {$i$ in 1, 2, ..., 10}
		\STATE tune $filt_{vol}~\sigma$ params $\rightarrow NET_{vol}(vol_{backup})$
		\STATE $vol_{filt} = filt_{vol}(vol)$
		\STATE $vol_{backup} = vol_{filt}$
		\ENDFOR
	\end{algorithmic}
\end{algorithm}

 The second reason is due to computational effeciency. While it is true that the noise is not stationary across the pixel elements of the projection view, tuning the parameters pixelwise would require more computational power than is currently available in most commercial CT systems due to the large size of the projection view. Additionally, tuning a single pixel value would require knowledge of how tuning that pixel would affect the image quality of the resultant volume, which implies that projection views cannot be tuned in isolation and would require information from other projection views. Therefore, we only estimate an optimal set of $\sigma$ values for each projection view, and not for each pixel within the projection view. 

To replace the human operator, we introduce convolutional neural networks, $NET_{sin}$ and $NET_{vol}$, for tuning the parameters in the projection and volume domains respectively. These CNNs take an image patch as an input, and output one of the two $\sigma$ parameters, and an action to tune the value of the chosen parameter. We allow the CNNs to take one of five possible actions (1) decrease the parameter value by 50\% (2) decrease the parameter value by 10\% (3) do not change the parameter value (4) increase the parameter value by 10\% (5) increase the parameter value by 50\%. To achieve the optimal parameter values, we allow 10 tuning steps. It is important to note that due to allowing multiple parameter tuning steps, the actual magnitudes of 10\% and 50\% chosen does not affect the final value too significantly.

\subsection{Training the Denoising Framework}

\begin{algorithm}
	\caption{Deep Q Network Training}
	\begin{algorithmic}[1]
		\REQUIRE $filt_{sin}, NET_{sin}, NET_{sin}', dataset, NET_{rew},$\par $filt_{vol}, NET_{vol}, NET_{vol}',$ 
		\FORALL {N = 1, 2, ......, 20}
		\FORALL {$vol_{noisy}$ in dataset}
		\STATE $proj = forward~project(vol_{noisy})$
		\STATE $proj_{backup} = proj$
		\FOR {$i$ in 1, 2, ..., 10}
		\STATE tune $filt_{sin}~\sigma$ params $\rightarrow NET_{sin}(proj_{backup})$
		\STATE $proj_{filt} = filt_{sin}(proj)$
		\STATE reward = $NET_{rew}(FDK(proj_{filt}))~-$\par
		$NET_{rew}(FDK(proj))$
		\STATE randomly sample proj for training:
		\STATE \hspace{\algorithmicindent}add (proj, $proj_{filt}$, reward, action) to pool
		\STATE $proj_{backup} = proj_{filt}$
		\STATE sample from pool:
		\STATE \hspace{\algorithmicindent}train $NET_{sin}$ using Eqn.\ref{lossEqn}
		\IF {$steps\%30~is~0$}
		\STATE $NET_{sin}' = NET_{sin}$
		\ENDIF
		\ENDFOR
		\STATE vol = FDK($proj_{filt}$)
		\STATE $vol_{backup}$ = vol
		\FOR {$i$ in 1, 2, ..., 10}
		\STATE tune$ filt_{vol}~\sigma$ params $\rightarrow NET_{vol}(vol_{backup})$
		\STATE $vol_{filt} = filt_{vol}(vol)$
		\STATE randomly sample vol for training:
		\STATE \hspace{\algorithmicindent} reward = $NET_{rew}(vol_{filt}) - NET_{rew}(vol)$
		\STATE \hspace{\algorithmicindent} add (vol, $vol_{filt}$, reward, action) to pool
		\STATE $vol_{backup} = vol_{filt}$
		\STATE sample from pool:
		\STATE \hspace{\algorithmicindent}train $NET_{vol}$ using~Eqn.\ref{lossEqn}
		\IF {$steps\%30~is~0$}
		\STATE $NET_{vol}' = NET_{vol}$
		\ENDIF
		\ENDFOR
		\ENDFOR
		\ENDFOR
	\end{algorithmic}
\end{algorithm}

\subsubsection{General Deep - Q Formulation}
The aim of the denoising scheme is to reduce noise, resulting in higher detectability of small and low contrast structures. To achieve this improved image quality, we task the scheme to learn a policy, which tunes the parameters to create an optimal image. The policy is learned via the Q - Learning approach, defined by the following equation:

\begin{equation}
	Q^*(s, a) = \underset{\pi}{max}[r^k + \gamma r^{k + 1} + \gamma^2 r^{k + 2}+...|s_k = s, a_k = a, \pi]
\end{equation}

\noindent where $Q^*$ is the optimal action value to be achieved, $\pi$ is the action choosing policy, $s$ is the current state, and $a$ is the action chosen at state $s$. We define $a$ as a function of parameter $p$ and tuning strength $t$. A property of $Q^*(s, a)$, as described by Bellman \cite{Bellman1966} is the following:

\begin{equation}
	\label{bellmanEqn}
	Q^*(s, a) = r + \gamma \cdot \underset{a'}{max} Q^*(s', a')
\end{equation}

\noindent where $r$ is the reward achieved by the optimal action $a$ at $s$. $s'$ is the state we observe when $a$ is taken at $s$. We parameterize the value action function with weights $W$, which can be determined by penalizing the deviation from the Bellman equation (Equation \ref{bellmanEqn}). This deviation can be mathematically represented by the following equation:

\begin{equation}
	L(W) = [r + \gamma \cdot max_{a'}Q(s', a'; W) - Q(s, a; W)]^2
\end{equation}

Following modern deep Q learning approaches \cite{Mnih2015} \cite{Hessel2018}, we introduce a new variable $W'$, representing an older version of the weights. We also introduce double deep - Q learning \cite{VanHasselt2016}, to prevent overestimations of our networks. We can then define our loss as:

\begin{equation}
	L(W) = [r + \gamma Q(s', Q\underset{a'}{max}(s', a'; W); W') - Q(s, a; W)]^2
\end{equation}

Both $NET_{sin}$ and $NET_{vol}$ have two paths, one to choose the parameter, and another to tune it. Therefore, we split the above loss function into two parts to train the networks. Our final loss function for each network is given as:

\begin{equation}
	\label{lossEqn}
	\begin{split}
		L(W) = [2r + \gamma Q(s', Q\underset{p'}{max}(s', p'); W); W') + \\\gamma Q(s', Q\underset{ t'}{max}(s', t'); W); W') - \\ (Q(s, p; W) + Q(s, t; W))]^2
	\end{split}
\end{equation}

We also introduce categorical learning \cite{Bellemare2017}, and a dueling architecture \cite{Wang2016} to improve the performance of our networks.

In standard Deep-Q learning, one of the major problems is that the future transitions are highly correlated to the current state, if the discount factor $\gamma$ is not set to zero. this would result in policies favoring a large number of steps. To remedy this, we introduce experience replay \cite{Mnih2015}, which stores each transition independently in a training pool. The training pools are sampled at random during the training step, resulting in each transition only being dependent on the current state.

\subsubsection{Network Architecture}

\begin{figure}
	\includegraphics[width=\linewidth]{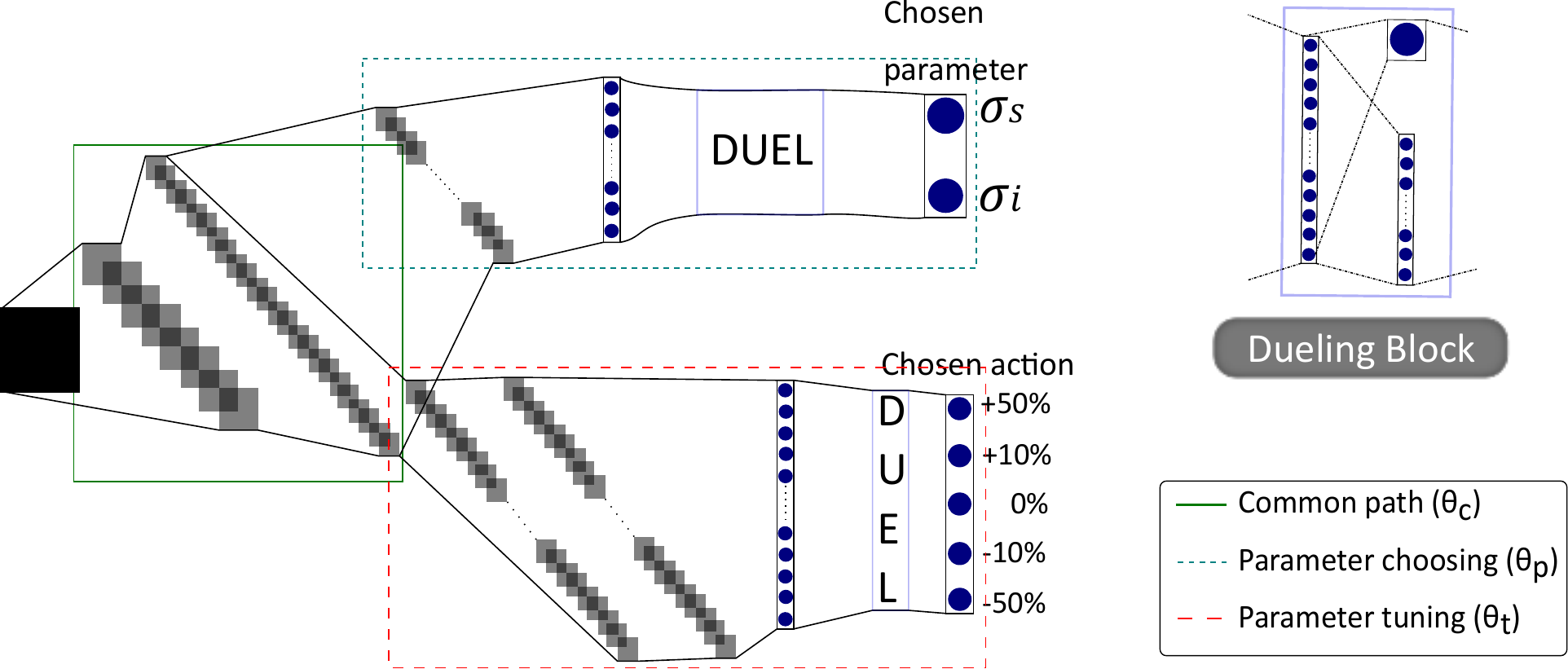}
	\caption{A general network architecture diagram for $NET_{sin}$ and $NET_{vol}$. Both networks have the same number of layers, although the number of filters and neurons differ. A dueling block (containing noisy layers) is shown in the top right. Both networks have three types of parameters, common parameters ($\theta_c$, green box), parameter choosing branch parameters ($\theta_p$, cyan box) and parameter tuning branch parameters ($\theta_t$, red box).}
	\label{tuningNet}
\end{figure}

We use a two branched architecture, with one branch used to select the parameter, and a second branch used to choose the tuning strength (Fig. \ref{tuningNet}). There are three types of parameters, $\theta_t$, which belong exclusively to the tuning branch, $\theta_p$, which belong exclusively to the parameter choosing branch, and $\theta_c$, which are shared among both branches.

The common parameters ($\theta_c$) of $NET_{sin}$ consists of two layers of 3 $\times$ 3 kernels, with 16 and 32 filters respectively. The parameter choosing branch ($\theta_p$) has a single 3 $\times$ 3 kernel with 32 filters, followed by a global average pooling layer, a fully connected layer with 64 neurons, and a dueling block. The parameter tuning branch ($\theta_t$) has two layers of 3 $\times$ 3 kernels, both with 64 filters, followed by a global average pooling layer, a fully connected layer with 128 neurons, and a dueling block. It would be possible to expand these networks to a larger number of parameters, by simply changing the size of the final output layer.

$NET_{vol}$ has a similar architecture to $NET_{sin}$ with some differences. Each of the network layers have double the number of filters or neurons compared to $NET_{sin}$. Global average pooling layers are replaced by flattening operations. The 3 $\times$ 3 kernels in $\theta_c$ are replaced with 3 $\times$ 3 $\times$ 3 kernels.

The dueling block contains three noisy layers. The first layer contains the same number of neurons as the previous layer. The second and third layers are used to predict the value and advantage respectively. Instead of directly regressing, we output a categorical distribution for each output. The support for the distributions ranges from 0 - 100 for $NET_{sin}$ and 0 - 200 for $NET_{vol}$. There are 51 bins.

Each layer in both networks is followed by a leaky ReLU activation layer. 

\subsubsection{Reward Network}
\label{rewNet}

\begin{figure}
	\centering
	\includegraphics[width=0.75\linewidth]{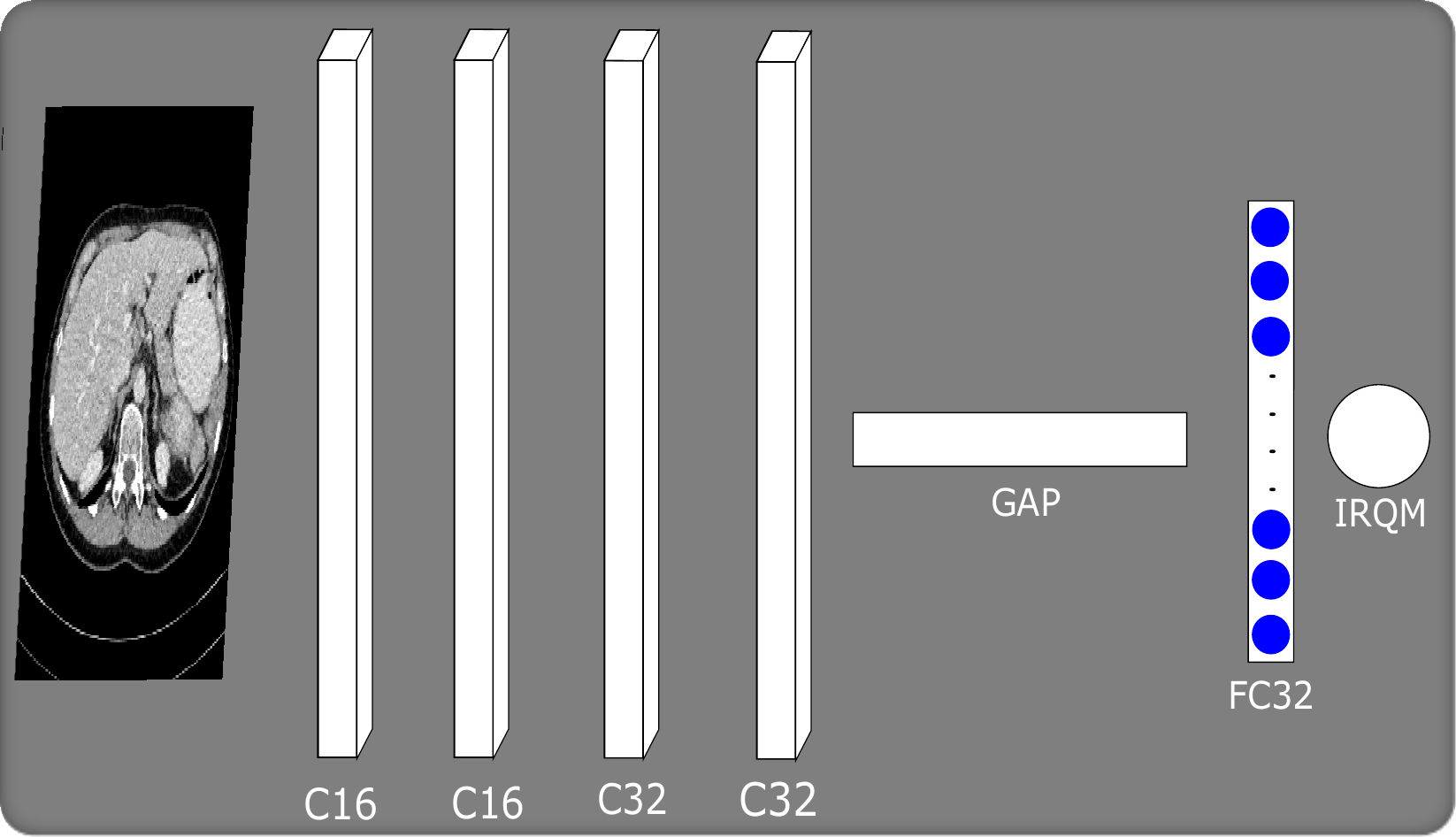}
	\caption{The network used as a reward function. C represents the number of filters in the convolutional layer, FC the number of neurons in the fully connected layer. GAP is a global average pooling operation, while IRQM is the output of the network.}
	\label{rewardNetIm}
\end{figure}

We create a small CNN to predict a quality score (Fig. \ref{rewardNetIm}). This CNN contains four convolutional layers with 3 $\times$ 3 kernels, followed by a global average pooling layer and a fully connected layer. The convolutional layers contain 16, 16, 32, and 32 filters. The fully connected layer contains 32 neurons. Each layer is followed by an eLU activation layer \cite{Clevert2016}, mimicing Patwari et al. \cite{Patwari2019}. The target for training is represented by the following equation:

\begin{equation}
	T(IM_1) = GSSIM(IM_1, IM_2) + \frac{1}{\frac{mean((IM_1 - IM_2)^2)}{ROI} + 1}
\end{equation}

\noindent where $IM_1$ is the noisy image, $IM_2$ is the clean image, and $GSSIM$ is the gradient structural similarity metric \cite{Chen2006}. The second term attempts to measure the noise power \cite{Kijewski1987} in the image. $ROI$ is the receptive field of the network (in this case 9 $\times$ 9).

This network is trained on a subset of five phantom images reconstructed at standard, 50\%, and 25\% dose. The dataset is augmented by flips, rotations, and Gaussian blurs.  Random patches of size 32, 48, 64, and 96 were extracted for training. Optimization was performed using the Adam optimizer\cite{Kingma2015} with a learning rate of $1 \times 10^{-4}$ over 30 epochs. By using phantom images, we ensure that no paired clinical training data are used in the training process at all. This training process occurs independently of the training of the parameter tuning networks.

The reward $r$ used in the Deep-Q scheme is given by the following equation:

\begin{equation}
	r = NET_{rew}(I(s')) - NET_{rew}(I(s))
\end{equation}

where $I(s)$ is the image at the current state and $I(s')$ is the image after applying the chosen action.

\subsubsection{Training Data and Hyperparameters}

We used the AAPM Low Dose CT Grand Challenge dataset \cite{Mccollough2016}, containing the body scans of 10 patients. During each epoch, we choose a random location on each patient, and simulate a cone - beam projection at the chosen location. We allow 10 tuning steps for tuning the parameters for $filt_{sin}$. We then denoise and reconstruct using FDK. We allow a further 10 tuning steps for $filt_{vol}$. 

The initial guesses for the $\sigma$ parameters are random integers between 1 and 25 for $filt_{vol}$ and random integers between 1 and 5 for $filt_{sin}$. The value of $\gamma$ is set to 0.99. The training is performed over 20 epochs over the whole training dataset. In each step, 2000 image blocks and 200 projections are added to the experience pools. 32 projections and 256 image blocks are chosen at random to train $NET_{sin}$ and $NET_{vol}$ respectively.  $W'$ is updated to $W$ every 30 steps. The Adam optimizer \cite{Kingma2015} was used for training with a learning rate of $1 \times 10^{-4}$. 

Our method was implemented in PyTorch 1.6 \cite{Paszke2019} and trained on a PC with a Titan Xp GPU and an Intel Xeon E5 - 2640 CPU. The ASTRA toolbox \cite{VanAarle2015,VanAarle2016} was used to simulate forward projections and reconstructions.

\subsection{Evaluation Data, Experimental Protocol and Metrics}

For evaluation, we use the freely available Mayo Clinic TCIA dataset \cite{Mccollough}. This dataset is available at https://doi.org/10.7937/9npb-2637. Since our method is trained on body CTs, we focus on the abdominal scans of 50 patients present in this dataset. We treat the standard dose volume as a ground truth volume and the 25\% dose volume as our noisy volumes. Non - anatomical parts of the image (eg. air and bed) were cropped out of the image, and analysis was conducted within the soft tissue window of [-160, 240] usually used for analysing abdominal scans.

\begin{figure}
	\includegraphics[width=\linewidth]{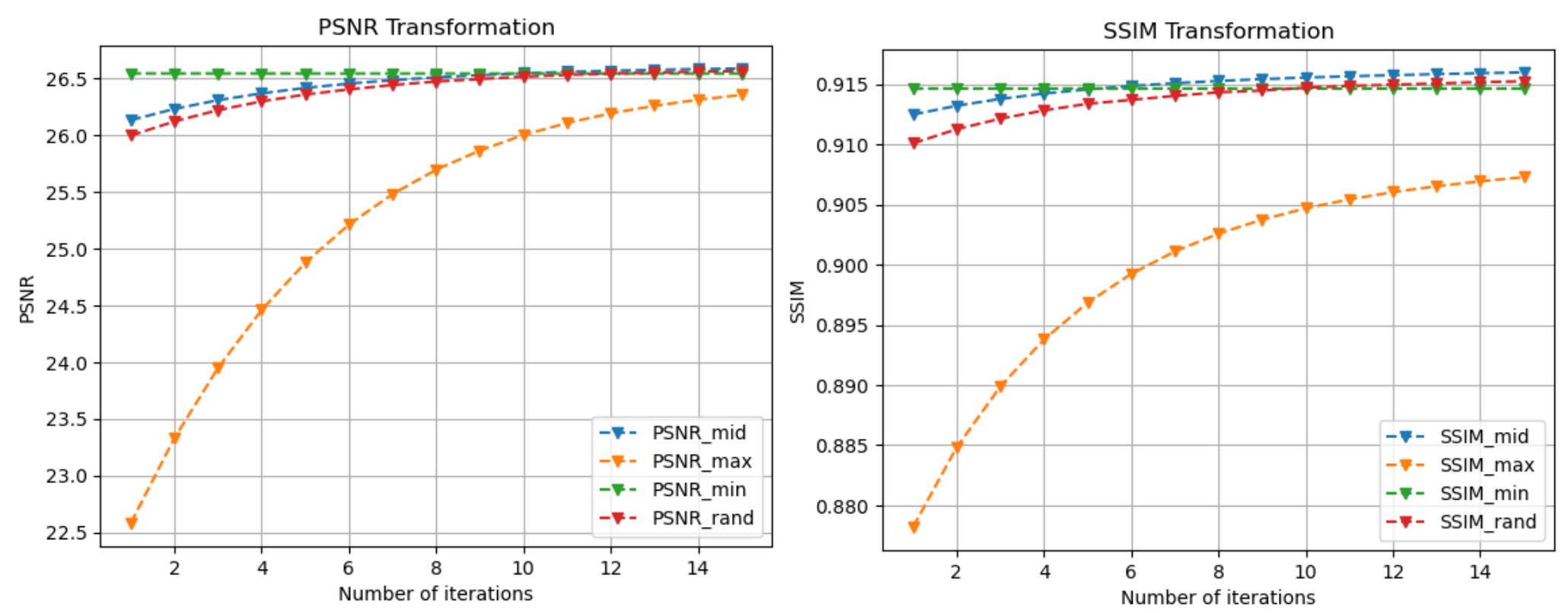}
	\caption{To illustrate convergence, we set the initial $\sigma$ values to the maximum ($\sigma_{filt}$=5, $\sigma_{vol}$=25), minimum ($\sigma_{filt}$=1, $\sigma_{vol}$=1), middle ($\sigma_{filt}$=3, $\sigma_{vol}$=13) and random values. We observe that the results converge to a similar values, given a sufficiently large number of steps. Since we test on abdominal scans, we stick with the middle initial values for all our subsequent experiments.}
	\label{convergenceIm}
\end{figure}

We analyze our results using two metrics: the peak signal to noise ratio (PSNR) and the structural similarity index (SSIM \cite{Wang2004}). The PSNR is a measure of noise suppression, and the SSIM is a metric of structural preservation. It is important to note, that in medical imaging, the SSIM is a significantly more important measures of image quality compared to the PSNR. Results are listed in Table \ref{resulttable} and will be elaborated upon in the following sections.

We use paired $t$-tests to compare the statistical significance of our metrics. We use a $p$-value of 0.05 as the threshold for statistical significance.

We applied our denoising framework onto simulated sequential scans of the given noisy volumes. The scans were simulated at 16 mm intervals. The source was placed 595 mm from the isocentre and 1085 mm from the detector. The simulated detector array was flat with 64 rows and 736 columns. The detector elements were of shape 1.285 mm $\times$ 1.094 mm. The knowledge of the scanner geometry was taken from the physical scanner information present in the accompanying DICOM-CT-PD files.

\section{Results}

\subsection{Framework Convergence}

During our training, we set the initial guesses of the $\sigma$ parameters for $filt_{sin}$ to a random number between 1 and 5, and the initial guesses of the $\sigma$ parameters for $filt_{vol}$ to a random number between 1 and 25. This should result in convergence to a similar value during inference, regardless of the initial guess. We test convergence on patient L056 from our test dataset. We set the initial values to 4 possibilities: minimum ($\sigma_{filt}$=1, $\sigma_{vol}$=1), maximum ($\sigma_{filt}$=5, $\sigma_{vol}$=25), middle ($\sigma_{filt}$=3, $\sigma_{vol}$=13) and pixelwise/projectionwise random integers. We allowed our method to run for 15 steps. 

\begin{table}
	\caption{Deep neural networks used as reference comparisons. The total number of parameters are listed. The number of parameters used for denoising are listed separately. Our method (RLDN) uses the fewest parameters during inference. MSE, AL, and PL are the mean square error, perceptual loss, and adversarial loss respectively.}
	\label{paramtable}
	\begin{center}
		\renewcommand{\arraystretch}{1.3}
		\begin{tabular}{|| c | c | c | c | c | c ||}
			\hline
			& \textbf{Generator} & \textbf{Discriminator} & \textbf{Loss} & \textbf{Training} & \textbf{Inference}\\
			& \textbf{parameters} & \textbf{parameters} & \textbf{functions} & \textbf{epochs} & \textbf{parameters}\\
			\hline
			\hline
			GAN3D \cite{Wolterink2017} & 862,753 & 2,105,345 & AL + MSE & 100 & 862,753\\
			CPCE3D \cite{Shan2018} & 118,209 & 1,210,305 & AL + PL & 100 & 118,209\\
			WGAN - VGG \cite{Yang2018} & 56,097 & 1,210,305 & AL + PL & 100 & 56,097\\
			QAE \cite{Fan2019} & 49,818 & \ding{55}  & MSE & 30 & 49,818\\
			REDCNN \cite{Chen2017} & 206,689 & \ding{55}  & MSE & 30 & 206,689\\
			CNN10 \cite{Zhang2017a} & \textbf{24,513} & \ding{55}  & MSE & 30 & 24,513\\
			\textbf{RLDN} & 1,002,946 & \ding{55}  & DeepQ & 30 & \textbf{4} \\
			\hline
		\end{tabular}
	\end{center}
\end{table}

We find that the PSNR and SSIM values converge, with a maximum PSNR difference of 0.25 and SSIM difference of 0.01 at epoch 15 (See Fig. \ref{convergenceIm}). At epoch 10, the PSNR and SSIM differences are already quite low at 0.5 and 0.01 respectively. The best results are achieved by the middle initial guesses. To guarantee an acceptable image quality in a reasonable amount of time, we set our initial guesses to the middle values and fix a maximum of 10 iterations for our framework for all subsequent experiments.

\subsection{Quantitative Denoising Performance}

\subsubsection{Performance on Quantitative Metrics}

\begin{figure}
	\centering
	\includegraphics[width=0.5\linewidth]{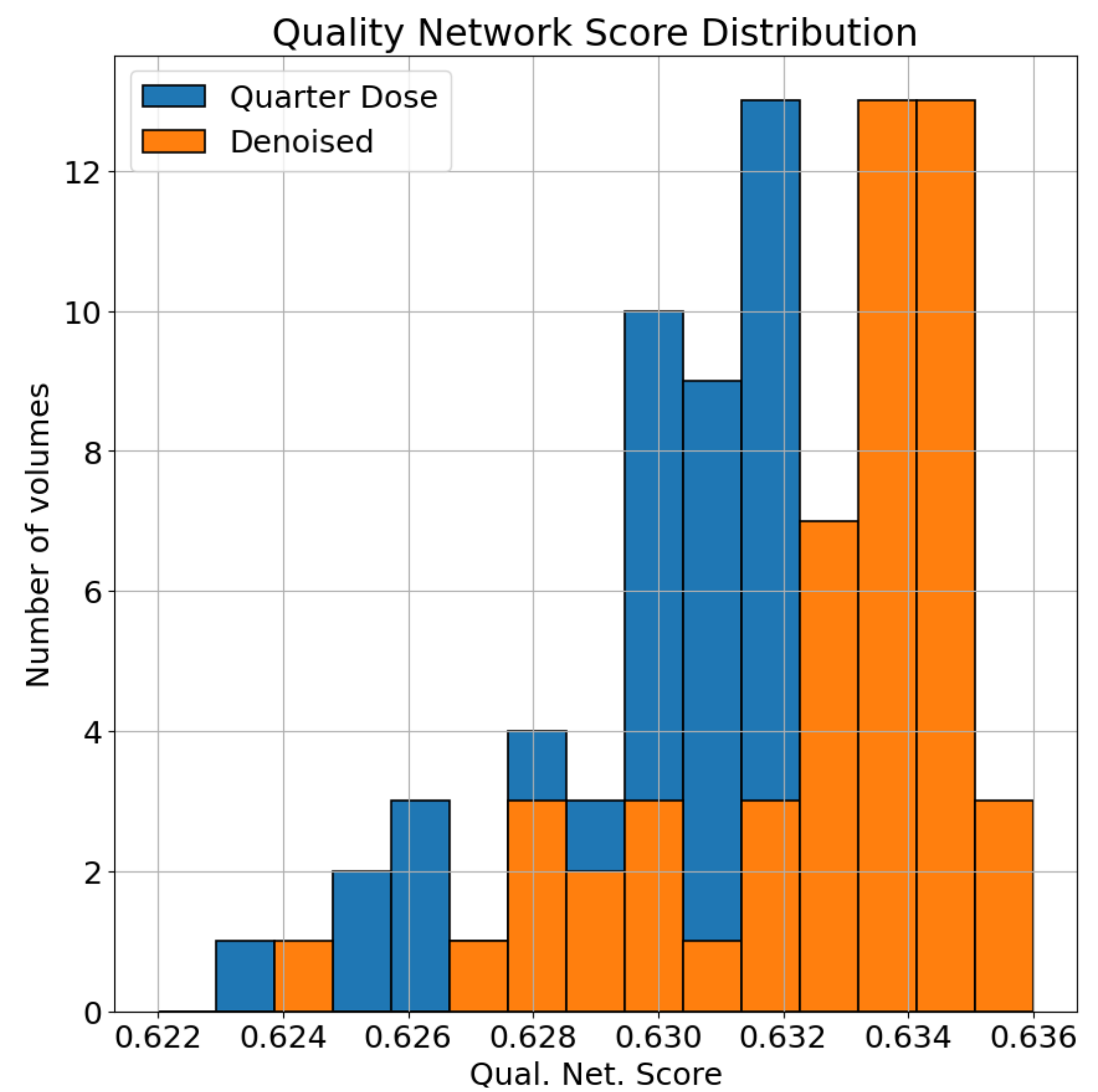}
	\caption{Change in the distribution of the estimated quality score before and after denoising. The peak of the distribution is shifted to the right, indicating a general increase in the quality score after denoising.}
	\label{irqmDist}
\end{figure}

After denoising, the mean PSNR was improved from 28.53 to 28.93 ($t$ = 2.834, $p$ = 0.0067)and the mean SSIM was increased from 0.8952 to 0.9204 ($t$ = 9.458, $p$ = 0.000). We observe that, the estimated quality score was increased after denoising. The peak of the distribution of the quality scores estimated by $NET_{rew}$ is shifted to the right (Fig. \ref{irqmDist}), indicating an increase in the estimated quality score after denoising.

\subsubsection{Comparison to Deep Neural Networks}

\begin{figure}
	\includegraphics[width=\linewidth]{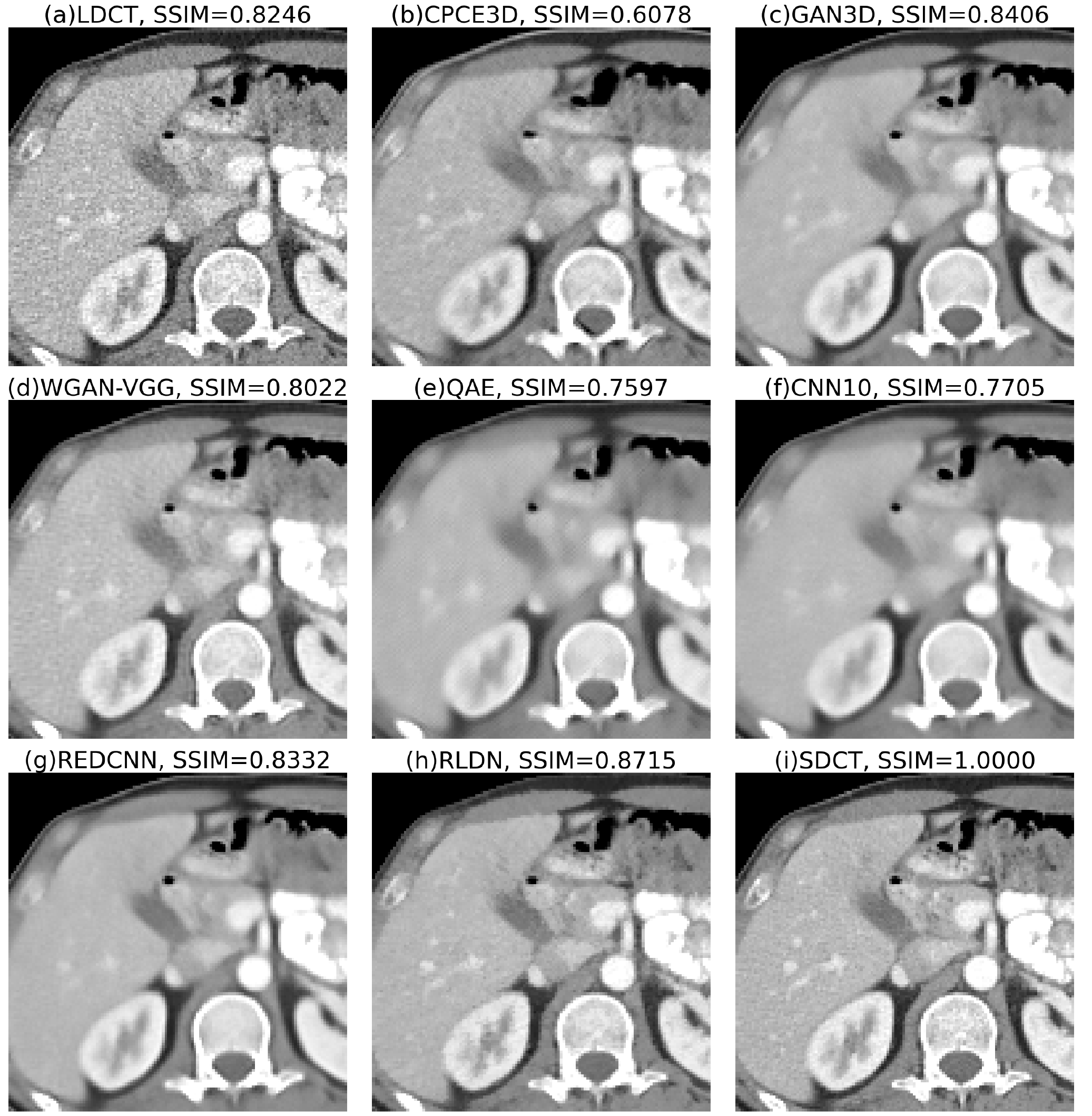}
	\caption{Comparison of the (a) low dose CT to the (b) CPCE3D (c) GAN3D, (d) WGAN - VGG (e) QAE, (f) CNN10, (g) REDCNN (h) our method (referred to as RLDN) and (i) standard dose CT. SSIM scores are shown in the captions. RLDN has the highest SSIM score (0.8715). Images are displayed with a window of [-160, 240].}
	\label{sotaCompIm}
\end{figure}

We compare our denoising framework to deep neural networks used for denoising (Fig. \ref{sotaCompIm}). In this study, we compare ourselves to GAN3D \cite{Wolterink2017}, CPCE3D \cite{Shan2018}, WGAN - VGG \cite{Yang2018}, QAE \cite{Fan2019}, REDCNN \cite{Chen2017} and CNN10 \cite{Zhang2017a}. To provide an fair ground for comparison, all the above networks were trained on the 10 patient volumes of the AAPM Grand Challenge dataset \cite{Mccollough2016}. The training hyperparameters, with the exception of the number of training epochs, were taken from the original papers. The number of training epochs and the loss functions used are present in Table \ref{paramtable}. The validation loss for the MSE loss based networks had stopped decreasing by the thirtieth epoch, therefore they were not trained for 100 epochs.

Our method has a superior PSNR and SSIM. Our method has 4 tunable parameters, whereas the networks mentioned in this section have several thousand trainable parameters (See Table \ref{paramtable}). In our experiments, methods using the MSE loss have fairly high PSNR values, but lower SSIM values (Table \ref{resulttable}). This is due to oversmoothing of images, which is visible in Figure \ref{sotaCompIm}, and is indicative of the MSE loss function. Using adversarial loss can improve performance (Table \ref{resulttable}), however, this has the tendency to generate some artifacts (Fig. \ref{sotaCompIm}) which may affect the quantitative results. Strong generated artifacts are visible in WGAN loss images (Fig. \ref{sotaCompIm}(b) and (d)).

Since we are comparing our method to image translation based denoising models, we also compare an ablation, which only tunes and filters in the volume domain (RLDN w/ only $filt_{vol}$) to our trained state-of-the-art models, as a fair comparison. We find that we still have a significantly higher PSNR and SSIM compared to state-of-the-art methods. This indicates, that even two tunable parameters may already be sufficient to exceed current state-of-the-art performance.

\subsection{Qualitative Denoising Performance}

While the results in Table \ref{resulttable} is promising for our method, it is important to evaluate our results in a clinical context. Therefore, we extract a region around a known tumor, with other visible structures, and observe the effect of all denoising methods on structure visibility. The image section is displayed in Fig. \ref{sotaClinicalIm}. We choose three structures to focus on, a known tumor (green arrow), a blood vessel (yellow arrow), and another high contrast structure (red arrow). To provide a comparison with standard non learnable denoising methods, we introduce BM3D as a reference method for this case.

\begin{table}
	\caption{Table containing the PSNR, SSIM, $p$-values scores for each of the reference networks and ablation studies on the abdomen scans in the TCIA dataset. The mean and standard deviation of each score are given. RLDN refers to our reinforcement learned denoiser. Paired $t$-tests were used to calculate the significant differences between RLDN and other results. The $p$-values of the PSNR and SSIM values when compared to RLDN are given in the table below.}
	\begin{center}
		\renewcommand{\arraystretch}{1.3}
		\begin{tabular}{ ||c || c | c || c | c || }
			\hline
			& \textbf{PSNR} & \textbf{SSIM} & \makecell{ \textbf{$p$-value} \\ \textbf{(PSNR)} } & \makecell{ \textbf{$p$-value} \\ \textbf{(SSIM)} }\\
			\hline
			Low Dose CT \cite{Mccollough} & 28.53 $\pm$ 2.204 & 0.8952 $\pm$ 0.0453 & 0.0067 & 0.000 \\ 
			\hline
			GAN3D \cite{Wolterink2017} & 28.63 $\pm$ 1.352 & 0.9047 $\pm$ 0.0269 & 0.000 & 0.000 \\  
			CPCE3D \cite{Shan2018} & 20.56 $\pm$ 0.5466 & 0.7544 $\pm$ 0.0351 & 0.000 & 0.000 \\
			WGAN - VGG \cite{Yang2018} & 25.99 $\pm$ 0.9579 & 0.8855 $\pm$ 0.0263 & 0.000 & 0.000 \\
			QAE \cite{Fan2019} & 26.12 $\pm$ 0.8584 & 0.8669 $\pm$ 0.0279 & 0.000 & 0.000 \\
			CNN10 \cite{Zhang2017a} & 26.95 $\pm$ 0.9811 & 0.8812 $\pm$ 0.0275 & 0.000 & 0.000 \\
			REDCNN \cite{Shen2018} & 28.58 $\pm$ 1.331 & 0.9085 $\pm$ 0.0264 & 0.000 & 0.000 \\
			\hline
			RLDN w/ only $filt_{sin}$ & 28.54 $\pm$ 1.492 & 0.9120 $\pm$ 0.0309 & 0.000 & 0.000\\
			RLDN w/ only $filt_{vol}$ & 28.93 $\pm$ 1.504 & 0.9205 $\pm$ 0.0279 & 0.7407 & 0.000\\
			RLDN w/ fixed filters & 28.36 $\pm$ 1.396 & 0.9162 $\pm$ 0.0278 & 0.000 & 0.000\\
			RLDN w/ fixed $filt_{sin}$ & 28.93 $\pm$ 1.505 & 0.9204 $\pm$ 0.0279 & 0.0002 & 0.2262\\
			RLDN w/ fixed $filt_{vol}$ & 28.28 $\pm$ 1.396 & 0.9150 $\pm$ 0.0275 & 0.000 & 0.000\\
			RLDN w/o $NET_{rew}$ & 28.68 $\pm$ 1.512 & 0.9153 $\pm$ 0.0308 & 0.000 & 0.000\\
			\hline
			\textbf{RLDN} & \textbf{28.93 $\pm$ 1.504} & \textbf{0.9204 $\pm$ 0.0279} & N/A & N/A \\
			\hline  	
		\end{tabular}
	\end{center}
	\label{resulttable}
\end{table}

We notice that BM3D, CNN10, REDCNN, and QAE blur the image significantly. While the tumor is still visible, the blood vessel and white structure are significantly blurred and cannot be clearly detected. GAN3D, while not blurring so aggressively, also reduces the contrast of the blood vessel and the white structure. WGAN - VGG and CPCE3D do not reduce the noise effectively, and the white structure is not clearly visible. Additionally, WGAN - VGG has the highest deviation from the CT numbers of the SDCT image. Our method reduces noise, but still allows the three structures do be detected easily.

We isolate the homogenous ROIs (green circles) to see if the statistical values agree with our observation. We compute the mean and standard deviations of the ROIs for all cases (Table \ref{roitable}). We find that the methods which we characterize as blurring the image, result in very low standard deviations, whereas those which we stated did not remove noise effectively, have high standard deviations. Our method has a lower standard deviation than the SDCT image, indicating even lower noise content, while not having a significant blurring effect.

\subsection{Ablation Studies}

\subsubsection{Filtering in Projection and Volume Domain}

We evaluate whether denoising in both projection and volume domain is actually necessary for achieving optimal image quality. Therefore, we experiment with tuning and applying only $filt_{sin}$ followed by a reconstruction, and tuning and applying only $filt_{vol}$, without any processing in the projection domain (Fig. \ref{ablaCompIm}(b) - (c)). 

We find that tuning and applying only $filt_{sin}$ results in significantly lower PSNR ($t$ = 19.81, $p$ = 0.000) and SSIM ($t$ = 14.36, $p$ = 0.000). Exclusively using and tuning $filt_{vol}$ does not signficantly affect the PSNR ($t$ = 0.3328, $p$ = 0.7407), however, it slightly increases the SSIM ($t$ = 4.984, $p$ = 0.000).

\subsubsection{Effect of Parameter Tuning}

It is important to determine, whether the improvement in image quality is due to the parameter tuning or simply due to the filtering action. To test this, we keep the values in $filt_{sin}$ fixed at the initial guess ($\sigma$s = 3), then we keep $filt_{vol}$ fixed at the initial guess ($\sigma$s = 13), and finally, we keep both filters fixed at their respective initial guesses. 

\begin{minipage}{\linewidth}
	\includegraphics[width=\linewidth]{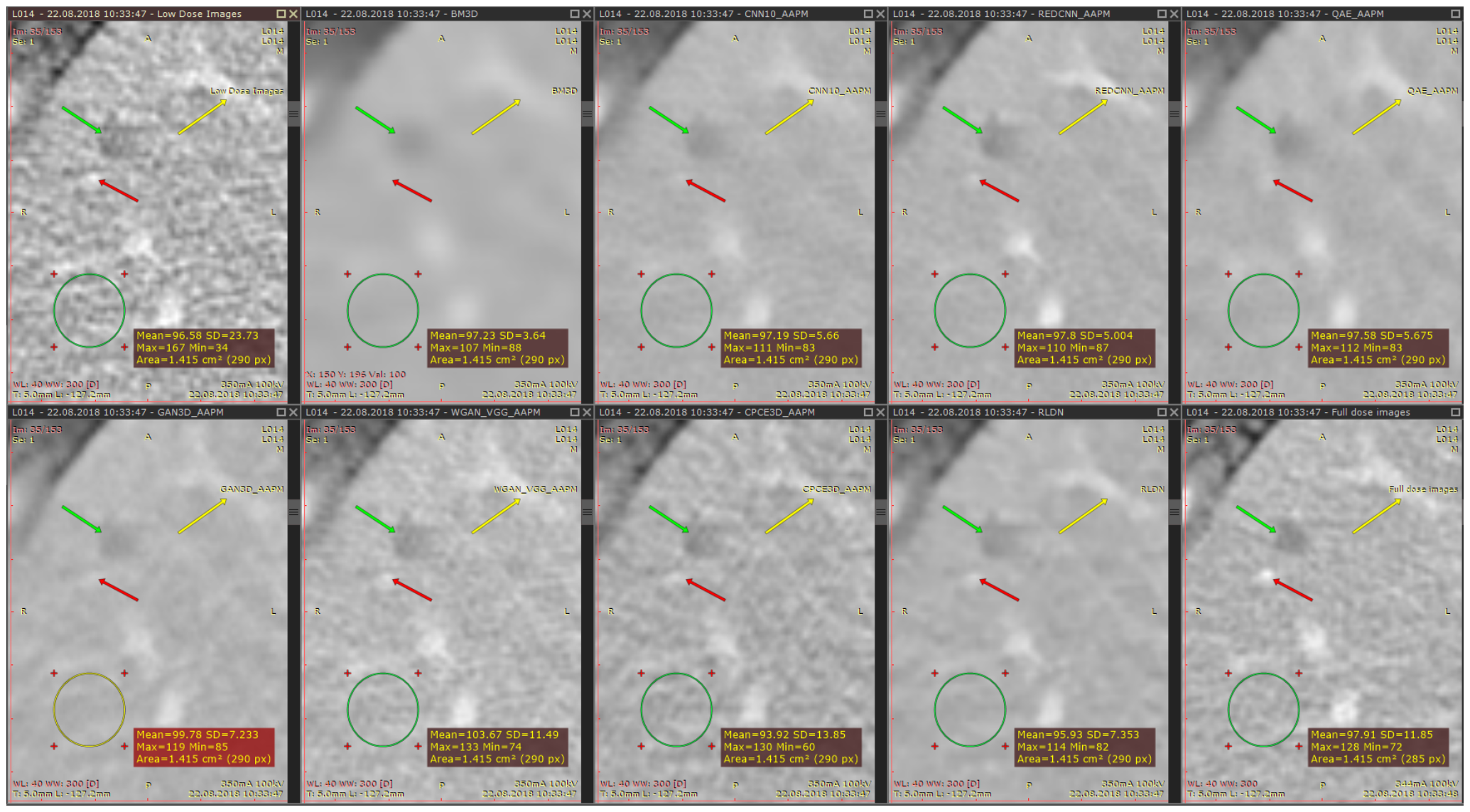}
	\captionof{figure}{Visualization of the area around a tumor (green arrow), containing a blood vessel (yellow arrow) and another visible structure (red arrow). We visualize the effect using (top row, from left to right) an LDCT image, denoised with BM3D, CNN10, REDCNN, QAE, (bottom row, from left to right) GAN3D, WGAN - VGG, CPCE3D, RLDN, and an SDCT image. A homogenous ROI (green circle) is chosen to obsrve the effect of each denoiser (see Table \ref{roitable}). Images are displayed with a window of [-110, 190].}
	\label{sotaClinicalIm}

	\captionof{table}{The mean and standard deviation of the homogenous ROI (marked by the green circle) in Fig. \ref{sotaClinicalIm} are shown here. The lowest standard deviation i.e. strongest denoising effect is caused by BM3D. The highest deviation from the mean is shown by WGAN - VGG.}
	\label{roitable}
	\begin{center}
		\renewcommand{\arraystretch}{1.3}
		\begin{tabular}{||c||c|c|c|c|c|c||}
			\hline
			& LDCT  & BM3D  & GAN3D & CPCE3D & WGAN-VGG \\ 
			\hline
			\textbf{Mean}  & 96.58 & 97.23 & 99.78 & 93.92  & 103.67                      \\ 
			\textbf{Standard Deviation} & 23.73 & \textbf{3.64}  & 7.23  & 13.85  & 11.49  \\ 
			\hline
			&  QAE   & REDCNN & CNN10 & \textbf{RLDN} & SDCT\\ 
			\hline
			\textbf{Mean} & 97.58 & 97.80  & 97.19 & 95.93  & 97.91                      \\ 
			\textbf{Standard Deviation} & 5.68  & 5.00 & 5.66  & 7.35 & 11.85 \\ 
			
			\hline
		\end{tabular}
	\end{center}
\end{minipage}

\newpage
We find that keeping the filters fixed results in a significant decrease in the PSNR ($t$ = 22.84, $p$ = 0.000) and SSIM ($t$ = 34.41, $p$ = 0.000). Keeping $filt_{sin}$ fixed resulted in a slight improvement in PSNR ($t$ = 4.021, $p$ = 0.0002), without any significant change in the SSIM ($t$ = 3.848, $p$ = 0.2262). Keeping $filt_{vol}$ fixed resulted in a significant decrease in PSNR ($t$ = 21.78, $p$ = 0.000) and SSIM ($t$ = 30.70, $p$ = 0.000) (see Fig. \ref{ablaCompIm} (d) - (f)). 

\subsubsection{Impact of the Reward Network}

In this study, we have used a reward network (Section \ref{rewNet}) to direct the reinforcement learning scheme. However, Shen et. al. \cite{Shen2018} used an objective reward function, which was defined by the following function:

\begin{equation}
r^k(x) = \frac{|S_{f^*}(x)|}{|S_{f^{k + 1}}(x) - S_{f^*}(x)|} - \frac{|S_{f^*}(x)|}{|S_{f^{k}}(x) - S_{f^*}(x)|}
\end{equation}

\noindent where $r^k$ is the reward at step $k$,$ S_{f^{k}}(x)$ is the image at step $k$, $S_{f^{k + 1}}(x)$ is the image after step $k$, and $S_{f^{*}}(x)$ is the ground truth image. We conduct a study by using this reward function. In this case, we use the standard dose volumes, which are part of the AAPM Grand Challenge dataset, as the ground truth volumes.

\begin{figure}
	\includegraphics[width=\linewidth]{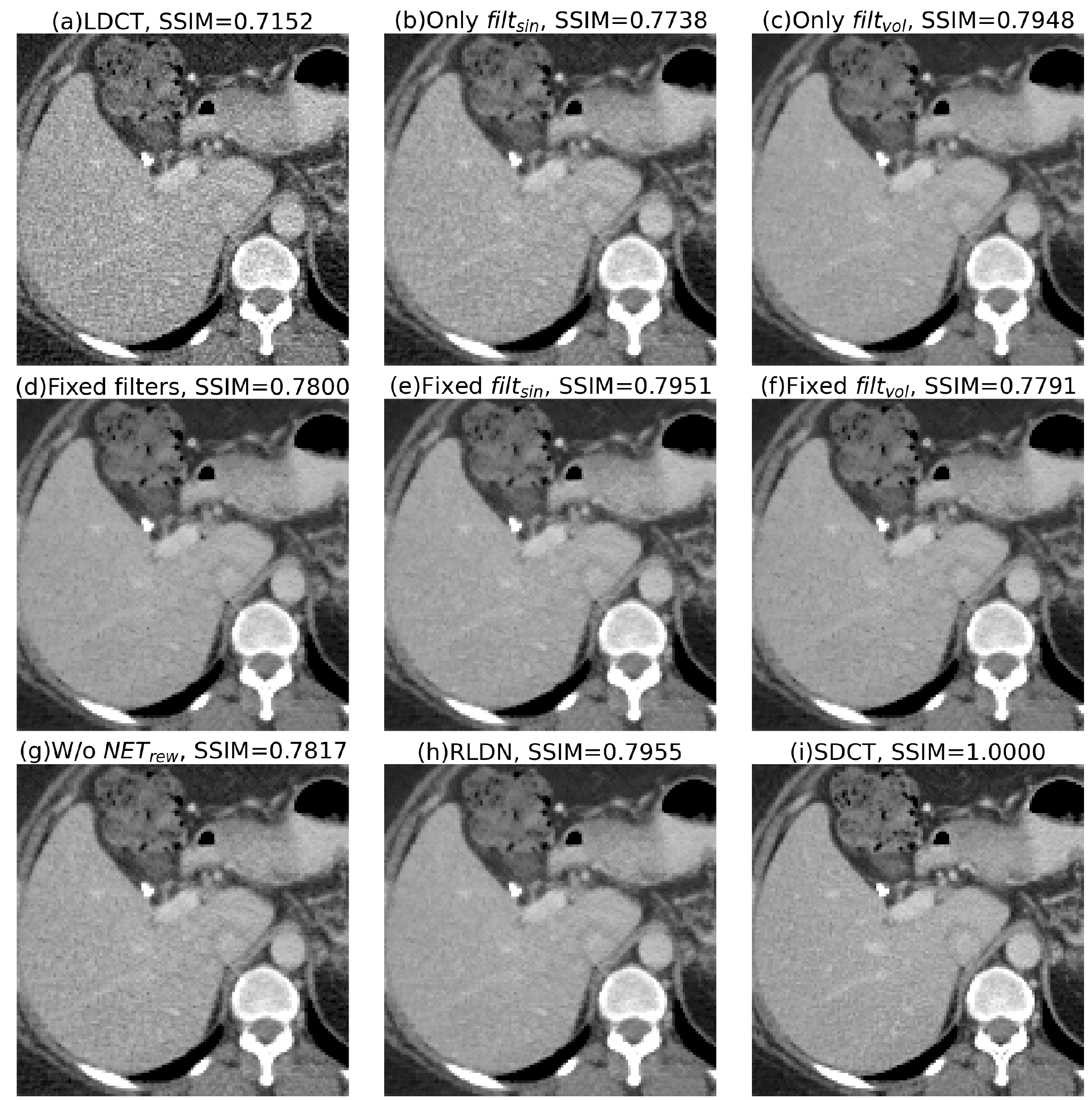}
	\caption{Comparison of (a) low dose CT to (b) filtering only in the projection domain (c) filtering only in the volume domain (d) filtering without parameter tuning (e) filtering without any parameter tuning in the projection domain (f) filtering without any parameter tuning in the volume domain (g) filtering without using a reward network during training (h) our method (represented by RLDN) and (i) the standard dose CT. SSIM scores are in the captions. RLDN achieves the highest SSIM score (0.7955). Images are displayed with a window of [-160, 240].}
	\label{ablaCompIm}
\end{figure}

We find that use of an objective reward function results in a significant decrease in PSNR ($t$ = 18.36, $p$ = 0.000) and SSIM ($t$ = 9.686, $p$ = 0.000) (see Fig. \ref{ablaCompIm} (g)).

\section{Discussion}

In this study, we developed a denoising approach which leveraged the physics of the CT problem for effective noise removal. We did this by implementing bilateral filtering in both the projection domain and the volume domain. Our method performs comparably and even outperforms several deep neural networks, with a large number of trainable parameters (Table \ref{paramtable}). We validated our method on an open - source dataset of 50 patients, which aids in reproducibility. The dataset we used for validation remains the only open source dataset for low dose CT scans, therefore generalizing across datasets was not possible.

The main advantage of our method lies in the simplicity and easy interpretability of the bilateral filter. The edge preserving denoising and low number of hand-tuned parameters made the bilateral filter the ideal choice for our denoising filter. Our method also has the advantage of not requiring large amounts of training data. 10 CT volumes were enough for our method to achieve competitive performance. Contrary to many reinforcement learning problems, our method only required 20 epochs of training, which was completed in less than 7 hours. This may however be due to the relatively straightforward nature of CT noise. Indeed, Shen et. al. \cite{Shen2018} also required less than 24 hours to complete training. A final important advantage of our method is the number of parameters. Only 4 tunable parameters are used in denoising the image (Table \ref{paramtable}). This is several orders of magnitude lower than GAN3D \cite{Wolterink2017}, WGAN-VGG \cite{Yang2018} and QAE \cite{Fan2019}.

%\begin{figure}
%	\includegraphics[width=\linewidth]{rlParamEffects}
%	\caption{Visualising the tuned maps for (a) low dose CT, (b) $\sigma_i$ (c) $\sigma_s$ and finally (d) the denoised result. In this image, tuning $\sigma_s$ plays a major role, whereas tuning $\sigma_i$ plays a lesser role. All parameter maps are only for $filt_{vol}$. Parameters are tuned based on CT values, however, images are displayed with a window of [-100, 200].}
%	\label{paramMap}
%\end{figure}

One of the key disadvantages of our method is the amount of time taken. Since a patch around each pixel must be passed through a network, this process can be time consuming. Denoising all 50 patient volumes took over 7 hours, as opposed to less than 15 minutes for our deep networks. Another reason for the excessive time is the resource heavy forward projection and FDK reconstruction. This could be remedied by the use of helical CT and reconstruction with WFBP or similar methods to remove the need for a forward projection, however, we have not yet implemented a reconstruction method for the DICOM-CT-PD format. The use of FDK also results in possible reconstruction artifacts. Although we did not notice any major artifacts, the possibility still exists. It is important to note that $NET_{sin}$, although trained on simulated projections, could be directly applied to acquired projections without retraining. This means that our framework could directly be applied to helical cone - beam scans when an adequate reconstruction technique is available.

Our results showed good denoising performance, with our method scoring the highest PSNR and SSIM vlaues, compared to state-of-the-art methods. It could be argued that the use of 10 patient training volumes is already significantly higher than the amount used in the original implementations of the state-of-the-art methods, however, even providing this additional data results in inferior denoising performance compared to our method. The ablations with fixed or no filtering in the projection domain resulted in slight improvements to the PSNR and SSIM. This can be because $NET_{sin}$ considers projections individually, and does not take into account other projections. Expanding $NET_{sin}$ into a 2.5D or 3D network may help to better tune and filter in the projection domain.

\section{Conclusions}
In this study, we investigated the development of a limited parameter noise removal technique, focusing on bilateral filtering and parameter tuning. Deep reinforcement learning was employed for the purposes of parameter tuning. We tested our method on a large publicly available dataset of abdominal CT scans. Our method achieved comparable performance to and in some cases even outperformed deep neural networks, which have a several orders of magnitude higher number of denoising parameters. We use only 4 tunable parameters for denoising. Training our agents works even without clean ground truth volumes, and is successful with only 10 volumes of training data. Both the tuning and filtering in the projection and the volume domain were shown to affect the quality of the reconstructed image. However, our method was extremely slow. Performance on different geometries and body parts was also not investigated. Future work on our method would be focused on generalizing to other body parts and scan geometries, and on speeding up our method.

\section*{Acknowledgements}
The authors would like to thank Mayo Clinic, for providing the TCIA dataset as an open source dataset.

\section*{Funding}
Funding for this work was provided by Siemens Healthineers AG.

\section*{Conflicts of Interest}
M. Patwari, R. Gutjahr, and R. Raupach are employees at Siemens Healthineers AG. A. Maier has received a research grant from Siemens Healthineers AG.

% following only if there is an appendix
%\section*{Appendix}
%\addcontentsline{toc}{section}{\numberline{}Appendix}
%Appendix text goes here if needed.

\section*{References}
\addcontentsline{toc}{section}{\numberline{}References}
\vspace*{-20mm}

% Following assumes you are using bibtex. However, for submission to the
% journal you MUST explicitly INCLUDE THE REFERENCES IN THE TEX FILE. 
% In that case you need the following

% \begin{thebibliography}{10}
% insert the .bbl file generated by bibtex here
	%This will be a series of entries from your .bib file formatted
	%something like
	%\bibitem{Me09}
        %{I.~Meijsing, B.~W.~Raaymakers, A.~J.~E.~Raaijmakers \it et al.},
        %\newblock {Dosimetry for the MRI accelerator: the impact of a 
	%magnetic field on the response of a Farmer NE2571 ionization chamber},
        %\newblock Phys. Med. Biol. {\bf 54}, 2993 -- 3002 (2009).

% \end{thebibliography}

% The following is when using bibtex and picks up the example.bib file

%\bibliography{Explicit address of .bib file}
\bibliography{./Bibliography}      %example.bib is on the same directory
% above points to where we find the master reference list
% and also causes the bibliography to be printed

% When creating your bibliography you should run bibtex on your local
% computer after running pdflatex on your .tex file. bibtex will
% generate a .bbl file.
% Copy the contents of this .bbl file into your main latex document,
% replacing the "\bibliography" command which was pointing at your .bib file.

% following defines style of .bbl file 

%\bibliographystyle{explicit relative path to medphy.bst}
\bibliographystyle{medphy}    %if this is installed on your system,
				    %it is not essential to have the    ./

% Note that you need to typeset once, then run bibtex, then typeset another
% two times to get the references working properly.

\end{document}